# Linear magnetoelectric effect in the antiferromagnetic $Sm_2BaCuO_5$

Premakumar Yanda[1], N. V. Ter-Oganessian[2] and A. Sundaresan[1*]

[1]*School of Advanced Materials and Chemistry and Physics of Materials Unit, Jawaharlal Nehru Centre for Advanced Scientific Research, Jakkur P.O., 560064, India*

[2]*Institute of Physics, Southern Federal University, Rostov-on-Don 344090, Russia*

We report the discovery of linear magnetoelectric effect in the well-known green phase compound, $Sm_2BaCuO_5$, which crystallizes in the centrosymmetric orthorhombic (*Pnma*) structure. Magnetization and specific heat measurements reveal the long-range antiferromagnetic ordering of $Cu^{2+}$ and $Sm^{3+}$-ions moments at $T_{N1}$ = 23 K and $T_{N2}$ = 5 K, respectively. Applied magnetic field induces dielectric anomaly at $T_{N1}$ whose magnitude increases with field, which results in significant (1.7%) magnetocapacitance effect. On the other hand, the dielectric anomaly observed in zero-applied magnetic field at $T_{N2}$ shows a small (0.4%) magnetocapacitance effect. Interestingly, applied magnetic field induces an electric polarization below $T_{N1}$ and the polarization varies linearly up to the maximum applied field of 9 T with the magnetoelectric coefficient α ~ 4.4 ps/m, demonstrating high magnetoelectric coupling. Below $T_{N2}$, the electric polarization decreases from 35 to 29 $\mu C/m^2$ at 2 K and 9 T due to ordering of Sm-sublattice. The observed linear magnetoelectricity in $Sm_2BaCuO_5$ is explained using symmetry analysis.

## I. INTRODUCTION

Magnetoelectric effect allows the control of electric polarization by magnetic field or magnetization by electric field, which is promising for applications in spintronic devices, magnetic field sensors, non-volatile memories, etc.[1–8] Materials showing this effect are known to be linear magnetoelectrics or multiferroics. In linear magnetoelectric materials, the induced electric polarization or magnetization is linearly proportional to the applied magnetic or electric field, respectively, which can be shown in the form $P = \alpha H$ or $M = \alpha E$, where α is the magnetoelectric coefficient.[1,8] However, these materials are restricted by symmetry requirements, involving simultaneous breaking of time reversal and spatial inversion symmetries.[2,9] This effect was first predicted in the antiferromagnetic $Cr_2O_3$ based on symmetry considerations by Dzyaloshinskii in 1959 and soon after confirmed experimentally by Astrov.[10,11] Since then there have been tremendous efforts to find new linear magnetoelectric materials.[12–16] Recently, linear magnetoelectric effect was reported in many materials such as $MnTiO_3$, $A_2M_4O_9$ (A = Nb & Ta, M = Mn, Fe & Co), $NdCrTiO_5$, $Cr_2WO_6$, $Co_3O_4$, $MnGa_2O_4$, $MnAl_2O_4$, $CoAl_2O_4$, etc.[17–25] Therefore, it is very challenging to find new materials within different structural types, which show magnetoelectric effect with strong coupling between magnetic and electric orders.

$Sm_2BaCuO_5$ belongs to well-known green phase $R_2BaCuO_5$ family of compounds, where $R$ stands for rare earth, which often appeared as impurity phases in the early synthesis of well-known 123 superconductors.[26,27] Later, they were used as pinning centres in $RBa_2Cu_3O_7$ superconductors to enhance the critical current density. These compounds crystallize in two different structural types depending on the size of the rare earth ion. The oxides with $R$-ions going from samarium to lutetium, including yttrium, crystallize in the orthorhombic structure (green phase) with space group *Pnma*, whereas the oxides with lanthanum, praseodymium, and neodymium, show tetragonal symmetry (brown phase) with space group *P4/mbm*.[27]

There are quite a few papers that report on the study of specific heat, spectral studies and magnetic properties of these compounds with orthorhombic structure revealing one or two magnetic phase transitions associated with copper and rare earth ions.[28–31] In the green phase compounds, the magnetic properties vary significantly for different *R*-ions demonstrating strong 4*f*-3*d* magnetic interactions. In view of rich variety of magnetic structures, these compounds would be interesting candidates for possible linear magnetoelectric or multiferroic properties.[32–34] To the best of our knowledge, there were no reports on the magnetoelectric properties of these compounds and, therefore, we have investigated the magnetoelectrical properties of $Sm_2BaCuO_5$. When we were about to communicate our results for publication, we become aware of a paper on ferroelectric and magnetoelectric properties of $R_2BaCuO_5$ (*R* = Er, Dy and Sm).[35]

In this article, we report the observation of linear magnetoelectric effect in the green phase oxide $Sm_2BaCuO_5$. In this compound, $Cu^{2+}$ ions order antiferromagnetically at $T_{N1}$ = 23 K, where electric polarization appears under applied magnetic fields and varies linearly with field. The origin of linear magnetoelectric effect has been discussed based on symmetry analysis.

## II. EXPERIMENTAL

Polycrystalline samples of $Sm_2BaCuO_5$ were prepared by heating the stoichiometric mixture of high purity $Sm_2O_3$ (preheated), $BaCO_3$ and $CuO$ at 950˚C in the air. X-ray powder diffraction patterns were recorded using a PANalytical empyrean diffractometer with Cu Kα1 radiation. Magnetization measurements were performed by a superconducting quantum interference device magnetometer (MPMS, Quantum Design). The specific heat (*Cp*) was measured in the physical property measurement system (PPMS, Quantum Design). To measure the dielectric properties and pyrocurrent (electric polarization), silver paste was coated on both sides of the disc shaped sample of dimension 5 mm * 5 mm area and 0.362 mm thickness, while the temperature and magnetic fields were controlled by PPMS. The dielectric constant as a function of temperature under different magnetic fields was recorded using an Agilent E4980A LCR meter. Prior to pyroelectric current measurement, the sample was poled in presence of electric and magnetic fields while cooling across the $T_{N1}$ and then short circuited for 15 min to remove stray charges. The temperature dependence of pyrocurrent was measured with a Keithley 6517A electrometer and electric polarization was obtained by integrating the pyrocurrent with respect to time.

## III. RESULTS AND DISCUSSION

The Rietveld refinement of room temperature X-ray powder diffraction data of $Sm_2BaCuO_5$ confirms the orthorhombic structure with the space group *Pnma*. Trace amount (~1%) of $Sm_2CuO_4$ phase was present as minor phase which was included in the refinement. The refined XRD pattern is shown in figure 1. The detailed structural parameters are given in Table I. The crystal structure can be considered as built up from distorted monocapped trigonal prisms $SmO_7$, which share one triangular face forming $Sm_2O_{11}$ blocks. These $Sm_2O_{11}$ blocks then share edges to form a three-dimensional network which demarcates the cavities where $Ba^{2+}$ and $Cu^{2+}$ are located. Each barium ion is coordinated by 11 oxygen atoms, while unusual $CuO_5$ forms isolated distorted square pyramid as reported earlier.[27]

Figure 2(a) shows temperature dependence of dc magnetic susceptibility data χ(T) measured with an applied field of 0.1 kOe under field cooled condition. These data show two clear anomalies in χ(T) at 23 K and 5 K corresponding to antiferromagnetic ordering of Cu and Sm-moments, respectively. The long-range magnetic ordering is confirmed by the λ shape anomalies in specific heat $Cp$(T) data as seen in figure 2(b). Overall, these results are similar to those reported earlier.[28,31] The presence of two anomalies around 23 K and 5 K in $Cp$(T) data suggest the independent ordering of Cu and Sm-moments, respectively. However, it is possible that the local field created by Cu-ordering may induce partial ordering of Sm-moments at $T_{N1}$ but it requires neutron diffraction studies to confirm this possibility. A broad hump around 65 K is seen in the magnetization data but there is no corresponding anomaly in $Cp$(T). The origin of this hump is due to presence of superconducting impurity phase Sm123, which was previously observed in some of the green phase compounds.[36] We did not observe this impurity phase in our laboratory X-ray powder diffraction data. However, this anomaly cannot influence the main results of this compound. The magnetic field dependent magnetization *M(H)* at different temperatures is shown in inset of figure 2(a). The curves at low temperatures are in good agreement with the antiferromagnetic ordering of both Cu and Sm ions and room temperature data resemble the paramagnetic behaviour.

Temperature dependence of the dielectric constant measured for different fields and the corresponding dissipation factor are shown in figure 3(a, b). We did not observe any dielectric anomaly at the magnetic ordering temperature of $Cu^{2+}$ ions under zero applied magnetic field. However, applied magnetic field induces a dielectric anomaly whose magnitude increases with field as shown in figure 3. Correspondingly, the loss data also display a peak at $T_{N1}$ with applied magnetic field. This behaviour is typical of linear magnetoelectric effect. Hence, the dielectric behaviour at $T_{N1}$ = 23 K signifies the role of Cu-spin structure and the presence of strong coupling between the magnetic and electric properties of $Sm_2BaCuO_5$ and possible magnetoelectric effect. On the other hand, we observed a broad dielectric anomaly at the Sm-ordering temperature ($T_{N2}$) in zero field which is almost insensitive to applied magnetic field. This compound shows dielectric relaxation behaviour at higher temperatures which is shown in figure S1 in the supplemental material.[37]

To verify whether these dielectric anomalies are associated with field induced electric polarization, which is a requirement for linear magnetoelectric effect, we have performed pyrocurrent measurements under various applied magnetic fields and a poling electric field E = +8.28 kV/cm. After magnetoelectric poling, the current was measured in the presence of magnetic field. In zero magnetic field, we did not observe any pyrocurrent peak (no polarization) at the magnetic ordering temperatures but there is a broad peak due to leakage current of 0.6 pA centered around 15 K as shown in figure S2 in the supplemental material.[37] The intrinsic pyrocurrent peak appears only under magnetic field at $T_{N1}$ and its magnitude increases with increasing magnetic field. In contrast, the leakage contribution remains almost constant with applied magnetic fields. To find the actual magnetoelectric current, we have subtracted the pyrocurrent measured under zero magnetic field from those measured under different magnetic fields as shown in figure 3(c). The appearance of pyrocurrent under the magnetic fields demonstrate the strong magnetoelectric effect in $Sm_2BaCuO_5$. The spontaneous polarization obtained by integrating the pyrocurrent with respect to time is shown in figure 3(d). With increasing magnetic field, the polarization increases monotonously to a value of 32 $\mu C/m^2$ at 7 K for H = 9 T. It is worth pointing out the behaviour of pyrocurrent at the independent Sm ordering temperature. A pyrocurrent peak appears at $T_{N2}$ but opposite to the direction of the peak at Cu ordering temperature, indicating the suppression of polarization at 5 K as seen in figure 3(d). This is due to the effect of independent ordering of Sm magnetic sublattice. There are few possibilities for the decrease of polarization below $T_{N2}$. It is possible that there is an additional contribution to the polarization from Sm moments in the temperature range $T_{N2}<T<T_{N1}$, due to its induced ordering at $T_{N1}$, which changes below $T_{N2}$. Alternatively, the independent Sm-ordering is strong enough to alter the copper magnetic structure decreasing electric polarization induced by it or which induces its own contribution to polarization opposite to that due to the copper sublattice. However, neutron diffraction studies are required in order to determine the exact nature of magnetic phase transitions at $T_{N1}$ and $T_{N2}$ and the resulting magnetic structures of Sm and Cu-sublattices as a function of temperature.

At 10 K, the polarization increases linearly with the magnetic field, see figure 4(a), which demonstrates the linear magnetoelectric effect in $Sm_2BaCuO_5$. The calculated magnetoelectric coefficient α of $Sm_2BaCuO_5$ is ~ 4.4 ps/m which is larger than that reported for the conventional linear magnetoelectric material $Cr_2O_3$. In fact, this value is higher than many of the known magnetoelectrics for example, $NdCrTiO_5$ (0.51 ps/m), $MnTiO_3$ (2.6 ps/m), $Co_3O_4$ (2.6 ps/m), $MnGa_2O_4$ (0.17 ps/m) indicating the strong magnetoelectric coupling in $Sm_2BaCuO_5$.[17,21,22,24] The observed value is quite high even though our sample is polycrystalline and bigger value is expected for the single crystal. As on today, the highest α known material is $TbPO_4$ with the value of ~730 ps/m but at very low temperature of 2.38 K.[8] In addition to this, $Sm_2BaCuO_5$ exhibits magnetodielectric effect as large as 1.7 % at magnetic field of 7 T near to transition temperature as shown in figure 4(b).

To confirm further the magnetoelectric effect, we have carried out the switching of polarization and dc bias measurements.[37,38] As shown in figure 5(a), the sign of the pyrocurrent and polarization switches simultaneously with the direction of poling electric field. Moreover, we observed a strong dc bias signal with positive polarization and negative depolarization peaks under applied magnetic field at the copper ordering temperature, as shown in figure 5(b). The absence of dc bias signal around 15 K reveals the broad pyrocurrent is due to leakage contribution. Overall these observations confirm that this transition is associated with magnetoelectric effect. To explain the microscopic mechanism, which is responsible for ferroelectricity, we need to know the magnetic structure of this compound.

As samarium is a strong absorbent of neutrons, it is difficult to perform a reliable neutron diffraction measurement. Hence, here we present the possible reasons for the linear magnetoelectric effect by using symmetry analysis along with theoretical calculations.

The analysis of literature data on $R_2$BaCuO$_5$ shows that these compounds experience one or two magnetic phase transitions at low temperatures (below ~25 K) depending on the nature of rare earth and the strength of the interaction between the rare earth and Cu-sublattices.[28,31] Neutron diffraction data reveal that various magnetic ordering wave vectors are found in the green phases including, e.g., $\vec{k}=(0,\frac{1}{2},0)$ in $R$ = Dy, Ho, Er, $\vec{k}=(0,\frac{1}{2},\frac{1}{2})$ in $R$ = Yb, Y, and $\vec{k}=(0,0,\frac{1}{2})$ in $R$ = Gd.[32–34] Furthermore, incommensurate magnetic structure is found in Gd$_2$BaCuO$_5$,[34] whereas a $\vec{k}=0$ magnetic structure is found in the ground state of Dy$_2$BaCuO$_5$.[32] The variety of magnetic ordering wave vectors can be arguably explained by the presence of three different magnetic sublattices and by a multitude of exchange constants, because simple geometrical calculation reveals that within a distance of, e.g., 5 Å there exist up to eleven different exchange paths. Our experimental results unambiguously show that Sm$_2$BaCuO$_5$ is a linear magnetoelectric below $T_{N1}$, which, together with the temperature dependence of electric polarization around $T_{N1}$, supports the appearance of magnetic ordering with $\vec{k}=0$ below $T_{N1}$ because of the following. According to the phenomenological theory of phase transitions, linear magnetoelectric effect induced by a magnetic structure with nonzero wave vector $\vec{k} \neq 0$ would be treated by the terms of the form $\xi_i^n MP$ in the thermodynamic potential, where $\xi_i$ is a (generally multicomponent) order parameter describing the antiferromagnetic structure, $M$ is magnetization, and $P$ is electric polarization. Due to nonzero wave vector one has $n > 1$, whereas the product $\xi_i^n M$ should be of even power with respect to magnetic order parameters because of time reversal symmetry. This gives the minimal value $n = 3$. In this case, however, the electric polarization at constant magnetic field will be proportional to $P \sim (T_{N1} - T)^\gamma$ with $\gamma = \frac{n}{2} > 1$ below the magnetic phase transition temperature, which contradicts the experimentally observed value $\gamma \approx 0.5$. Thus, one can conclude that the appearing magnetic structure is described by $\vec{k} = 0$ and induces linear magnetoelectric effect due to interaction of the form $\xi MP$, which results in $P \sim (T_{N1} - T)^\gamma$ with $\gamma \approx 0.5$ below $T_{N1}$. It has to be noted that under applied magnetic field (i.e. when $M \neq 0$) the phase transition at $T_{N1}$ becomes a proper ferroelectric phase transition, because the order parameters $\xi$ and $P$ have the same symmetry if $M \neq 0$. This explains the magnetic field-induced dielectric anomaly at $T_{N1}$.

Given the absence of neutron diffraction data on $Sm_2BaCuO_5$ one can tentatively assume the same relative spin arrangement as found in the low temperature magnetic structure of $Dy_2BaCuO_5$. In the *Pnma* crystal structure of the green phase the copper ions as well as both inequivalent rare-earths are located in positions 4c with coordinates: 1 $\left(x, \frac{1}{4}, z\right)$, 2 $\left(\frac{1}{2} - x, \frac{3}{4}, \frac{1}{2} + z\right)$, 3 $\left(-x, \frac{3}{4}, -z\right)$, and 4 $\left(\frac{1}{2} + x, \frac{1}{4}, \frac{1}{2} - z\right)$. For each magnetic sublattice one can define the basis vectors $\vec{F} = \vec{S}_1 + \vec{S}_2 + \vec{S}_3 + \vec{S}_4$, $\vec{G} = \vec{S}_1 - \vec{S}_2 + \vec{S}_3 - \vec{S}_4$, $\vec{C} = \vec{S}_1 + \vec{S}_2 - \vec{S}_3 - \vec{S}_4$, and $\vec{A} = \vec{S}_1 - \vec{S}_2 - \vec{S}_3 + \vec{S}_4$, where $\vec{S}_i$ is the spin of atom $i$. Thus, $\vec{F}$ is a ferromagnetic order parameter, whereas $\vec{G}$, $\vec{C}$, and $\vec{A}$ describe antiferromagnetic structures. In $Dy_2BaCuO_5$ the low temperature magnetic structure is described by the order parameters $C_x$ and $A_z$ transforming according to irreducible representation $\Gamma^{4-}$. It can be found that such relative spin arrangement breaks inversion symmetry, because the pairs of atoms 1 and 3, as well as 2 and 4, which are connected by spatial inversion, have oppositely directed magnetic moments. Thus, this magnetic structure allows linear magnetoelectric effect with magnetoelectric interactions of the form

$$C_xF_xP_z,$$

$$C_xF_zP_x,$$

$$A_zF_xP_z,$$

$$A_zF_zP_x.$$

In our measurements of electric polarization, we employed parallel $H \parallel E$ geometry, however similar results were obtained for $H \perp E$ due to the ceramic nature of the samples.

It has to be noted, that the same relative spin arrangement as in the low temperature phase of $Dy_2BaCuO_5$ is found in $Yb_2BaCoO_5$ below $T_N \approx 9.4$ K, which means that $Yb_2BaCoO_5$ should also experience linear magnetoelectric effect below this temperature.[39] Neutron diffraction experiments or single crystal magnetoelectric measurements in different geometries are required in order to confirm the suggested magnetic structure of $Sm_2BaCuO_5$. Alternative magnetic structures allowing linear magnetoelectric effect include those, which are described by inversion-odd irreducible representations in the Brillouin zone centre, i.e., $\Gamma^{1-}(A_xC_z)$, $\Gamma^{2-}(A_y)$, and $\Gamma^{3-}(C_y)$, however the symmetry analysis above will remain qualitatively the same.

As noted above, according to our experimental results $Sm_2BaCuO_5$ exhibits considerable magnetoelectric effect. From our point of view the high value of magnetoelectric coefficient can be related to the presence of rare earth ions, which introduce strong spin-lattice coupling due to high spin-orbit interaction. From the analysis of literature one can conclude that rare earth-containing magnetoelectrics generally show high magnetically induced electric polarization.[40] Furthermore, the green phase compounds often develop strongly non-collinear magnetic ordering with magnetic moments lying predominantly in the $ac$ plane.[32] In fact, both the $Cu^{2+}$ and $Sm^{3+}$ ions are located at positions with $\sigma_y$ symmetry (mirror plane perpendicular to the $b$ axis), which implies for all magnetic ions the existence of local electric dipole moments lying in the *ac* plane. Thus, the single-ion contribution to the magnetoelectric effect is allowed by symmetry and can be large for the rare earth ions, as is the case in rare earth manganites *R*MnO$_3$.[41] The strong influence of rare earth ions on magnetic field-induced electric polarization is further confirmed by strong dielectric anomaly at Sm$^{2+}$ ordering temperature $T_{N2}$ even in zero magnetic field.

Contrary to our results of linear magnetoelectric efffect in Sm$_2$BaCuO$_5$ at low temperature, the recent report on $R_2$BaCuO$_5$ (*R* = Er, Dy and Sm) claims that all the three compounds undergo ferroelectric transitions at high temperatures, ~235 K, ~232 K and ~184 K, respectively, which has been attributed to structural transition from non-polar (*Pnma*) to polar (*Pna*2$_1$) space group as inferred from synchrotron powder diffraction.[35] However, earlier neutron diffraction studies on $R_2$BaCuO$_5$ family of compounds strongly suggest that the structure remains non-polar (*Pnma*) down to the lowest temperature measured.[32–34] Further, the authors have attributed the symmetric and broad pyrocurrent peaks at high temperature to ferroelectricity. In the case of Sm$_2$BaCuO$_5$, we observe two peak-like features in the pyrocurrent data at a lower temperature, which shifts to high temperature with different warming rates, indicating the extrinsic origin of this peak as shown in figure S3(a) in supplemental material.[37] Furthermore, the extrinsic origin of polarization is confirmed by the dc bias measurement in which the pyrocurrent increases continuously as shown in figure S3(b), indicating the absence of ferroelectric behaviour.[37]

**IV. CONCLUSIONS**

We systematically investigated the linear magnetoelectric effect in the well-known green phase Sm$_2$BaCuO$_5$ by using magnetic, specific heat, dielectric and pyrocurrent measurements. This compound exhibits antiferromagnetic ordering at 23 K where we observed the appearance of electric polarization under applied magnetic field. Sm$_2$BaCuO$_5$ shows considerable linear magnetoelectric effect with strong coupling coefficient. Further single crystal studies are required to better understand the observed magnetoelectric effect.

**Acknowledgement**

Authors would like to acknowledge Sheikh Saqr Laboratory (SSL) and International Centre for Materials Science (ICMS) at Jawaharlal Nehru Centre for Advanced Scientific Research (JNCASR) for various experimental facilities. A. S. acknowledges BRICS Research Project, Department of Science Technology, Government of India for financial support. P. Y. acknowledges University Grants Commission (UGC) for Ph.D. fellowship (JNC/S0580). N.V.T. acknowledges financial support by the Russian Foundation for Basic Research grant No. 18-52-80028 (BRICS STI Framework Programme).

**References**

[1] M. Fiebig, J. Phys. D. Appl. Phys. **38**, R123 (2005).

[2] W. Eerenstein, N.D. Mathur, and J.F. Scott, Nature **442**, 759 (2006).

[3] T. Kimura, T. Goto, H. Shintani, K. Ishizaka, T. Arima, and Y. Tokura, Nature **426**, 55 (2003).

[4] D. Khomskii, Physics (College. Park. Md). **2**, 20 (2009).

[5] S.-W. Cheong and M. Mostovoy, Nat Mater **6**, 13 (2007).

[6] J.F. Scott, J. Mater. Chem. **22**, 4567 (2012).

[7] S. Fusil, V. Garcia, A. Barthélémy, and M. Bibes, Annu. Rev. Mater. Res. **44**, 91 (2014).

[8] J.-P. Rivera, Eur. Phys. J. B **71**, 299 (2009).

[9] R.E. Newnham, *Properties of Materials: Anisotropy, Symmetry, Structure* (Oxford University Press on Demand, 2005).

[10] I.E. Dzyaloshinskii, Sov. Phys. JETP **10**, 628 (1960).

[11] D.N. Astrov, Sov. Phys. JETP **11**, 708 (1960).

[12] G.T. Rado, Phys. Rev. Lett. **13**, 335 (1964).

[13] G.T. Rado, J.M. Ferrari, and W.G. Maisch, Phys. Rev. B **29**, 4041 (1984).

[14] M. Mercier and J. Gareyte, Solid State Commun. **5**, 139 (1967).

[15] I. Kornev, M. Bichurin, J.-P. Rivera, S. Gentil, H. Schmid, A.G.M. Jansen, and P. Wyder, Phys. Rev. B **62**, 12247 (2000).

[16] R.M. Hornreich, Solid State Commun. **7**, 1081 (1969).

[17] N. Mufti, G.R. Blake, M. Mostovoy, S. Riyadi, A.A. Nugroho, and T.T.M. Palstra, Phys. Rev. B - Condens. Matter Mater. Phys. **83**, 1 (2011).

[18] E. Fischer, G. Gorodetsky, and R.M. Hornreich, Solid State Commun. **10**, 1127 (1972).

[19] Y. Fang, W.P. Zhou, S.M. Yan, R. Bai, Z.H. Qian, Q.Y. Xu, D.H. Wang, and Y.W. Du, J. Appl. Phys. **117**, 17B712 (2015).

[20] A. Maignan and C. Martin, Phys. Rev. B **97**, 161106 (2018).

[21] Y. Fang, Y.Q. Song, W.P. Zhou, R. Zhao, R.J. Tang, H. Yang, L.Y. Lv, S.G. Yang, D.H. Wang, and Y.W. Du, Sci. Rep. **4**, 3860 (2014).

[22] J. Hwang, E.S. Choi, H.D. Zhou, J. Lu, and P. Schlottmann, Phys. Rev. B **85**, 24415 (2012).

[23] Y. Fang, L.Y. Wang, Y.Q. Song, T. Tang, D.H. Wang, and Y.W. Du, Appl. Phys. Lett. **104**, 132908 (2014).

[24] R. Saha, S. Ghara, E. Suard, D.H. Jang, K.H. Kim, N. V Ter-Oganessian, and A.

Sundaresan, Phys. Rev. B **94**, 14428 (2016).

[25] S. Ghara, N. V. Ter-Oganessian, and A. Sundaresan, Phys. Rev. B **95**, 1 (2017).

[26] C. Michel and B. Raveau, J. Solid State Chem. **43**, 73 (1982).

[27] A. Salinas-Sanchez, J.L. Garcia-Muñoz, J. Rodriguez-Carvajal, R. Saez-Puche, and J.L. Martinez, J. Solid State Chem. **100**, 201 (1992).

[28] V. V Moshchalkov, N.A. Samarin, I.O. Grishchenko, B. V Mill, and Z. J., Solid State Commun. **78**, 879 (1991).

[29] I. V Paukov, M.N. Popova, and B. V Mill, Phys. Lett. A **169**, 301 (1992).

[30] A. Salinas-Sánchez, R. Sáez-Puche, and M.A. Alario-Franco, J. Solid State Chem. **89**, 361 (1990).

[31] R.Z. Levitin, B. V Mill, V. V Moshchalkov, N.A. Samarin, V. V Snegirev, and J. Zoubkova, J. Magn. Magn. Mater. **90**, 536 (1990).

[32] I. V Golosovsky, V.P. Plakhtii, V.P. Kharchenkov, J. Zoubkova, B. V Mill, M. Bonnet, and E. Roudeau, Sov. Physics. Solid State **34**, 782 (1992).

[33] I. V Golosovsky, P. Böni, and P. Fischer, Solid State Commun. **87**, 1035 (1993).

[34] A.K. Ovsyanikov, I. V Golosovsky, I.A. Zobkalo, and I. Mirebeau, J. Magn. Magn. Mater. **353**, 71 (2014).

[35] A. Indra, S. Mukherjee, S. Majumdar, O. Gutowski, M. v. Zimmermann, and S. Giri, Phys. Rev. B **100**, 014413 (2019).

[36] L. Baum, Doctoral dissertation, Facultad de Ciencias Exactas (2003).

[37] See Supplemental Material at http://link.aps.org/supplemtal/ for dielectric relaxation, leakage contribution, high temperature pyrocurrent/dc bias signal and DC bais technique.

[38] C. De, S. Ghara, and A. Sundaresan, Solid State Commun. **205**, 61 (2015).

[39] J. Hernández-Velasco, R. Sáez-Puche, and J. Rodríguez-Carvajal, J. Alloys Compd. **275**, 651 (1998).

[40] T. Kimura, G. Lawes, T. Goto, Y. Tokura, and A.P. Ramirez, Phys. Rev. B **71**, 224425 (2005).

[41] V.P. Sakhnenko and N. V Ter-Oganessian, J. Phys. Condens. Matter **24**, 266002 (2012).

**Figure captions:**

Figure 1. Rietveld refined room temperature X-ray diffraction pattern of $Sm_2BaCuO_5$. The second row vertical tick marks indicate the secondary phase $Sm_2BaCuO_5$ (1%).

Figure 2. a) Magnetic susceptibility as a function of temperature measured with magnetic field of 0.1 kOe under field cooled sequence. Inset shows the magnetization curves against magnetic field at different temperatures. b) Magnetic susceptibility and specific heat in the low temperature region.

Figure 3. (a, b) Temperature dependent dielectric constant and loss factor measured under different magnetic fields with 50 kHz frequency. (c, d) Leakage current subtracted pyroelectric current as a function of temperature under different magnetic fields and poling electric field E = +8.28 kV/cm and corresponding polarization obtained by integrating pyrocurrent with respect to time.

Figure 4. a) Polarization as a function of magnetic field measured at 10 K. b) Magnetic field change in dielectric constant at 2, 15, 23, and 35 K measured under frequency of 50 kHz.

Figure 5. (a, b) Electric field switching of electric polarization and dc bias signal recorded at different magnetic fields under poling electric field E = 8.28 kV/cm.

**Table I.** Structural parameters of $Sm_2BaCuO_5$ obtained from Rietveld refinement. Space group: *Pnma*; a = 12.4140 (1) Å, b = 5.7647 (1) Å, c = 7.2798 (2) Å, Vol: 520.968 (6) Å$^3$; $\chi 2$ = 1.53; Bragg R-factor = 3.98 (%), Rf -factor = 4.15 (%).

| Atom | Site | x | y | z | $B_{iso}$ (Å$^2$) |
|---|---|---|---|---|---|
| Sm1 | 4c | 0.2886 (1) | 0.2500 | 0.1142 (2) | 0.033 (22) |
| Sm2 | 4c | 0.0737 (1) | 0.2500 | 0.3938 (2) | 0.033 (22) |
| Ba | 4c | 0.9062 (1) | 0.2500 | 0.9301 (2) | 0.253 (36) |
| Cu | 4c | 0.6584 (3) | 0.2500 | 0.7132 (5) | 0.082 (81) |
| O1 | 8d | 0.4342 (10) | -0.0124 (18) | 0.1746 (11) | 1.000 |
| O2 | 8d | 0.2271 (8) | 0.5120 (19) | 0.3505 (16) | 1.000 |
| O3 | 4c | 0.0951(12) | 0.250 | 0.0694 (19) | 1.000 |

Figure 1:

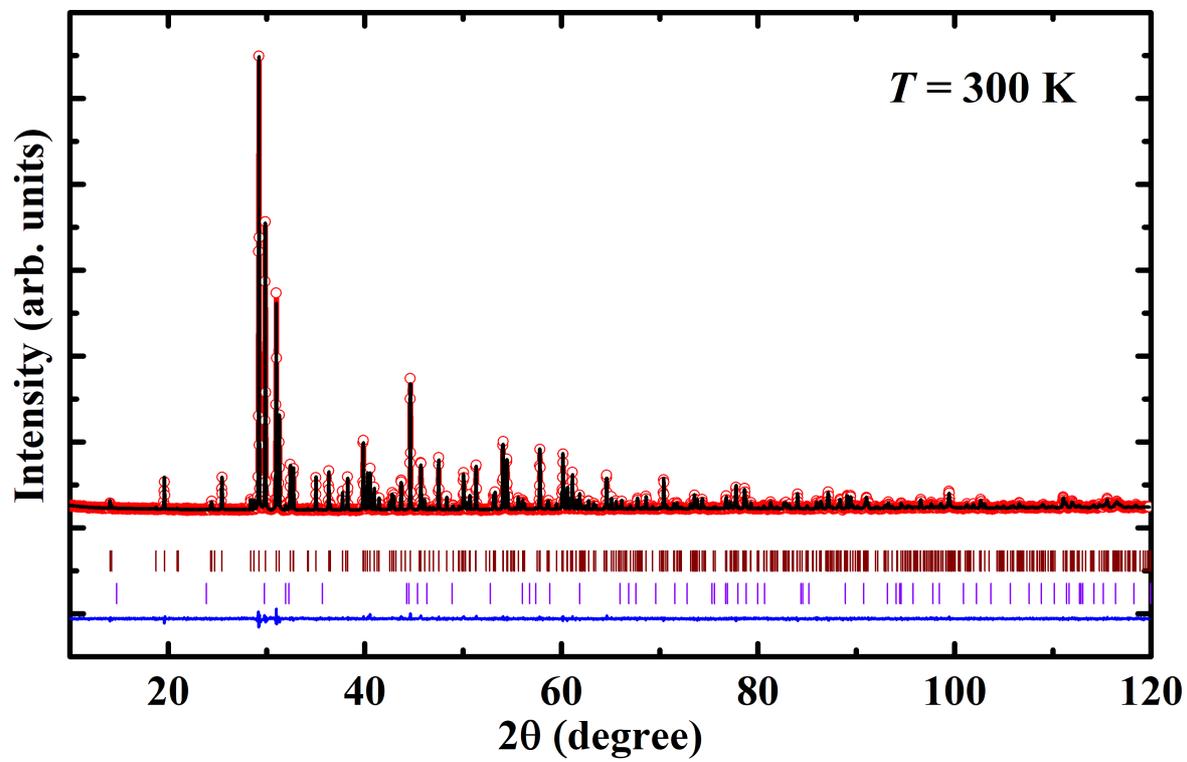

Figure 2:

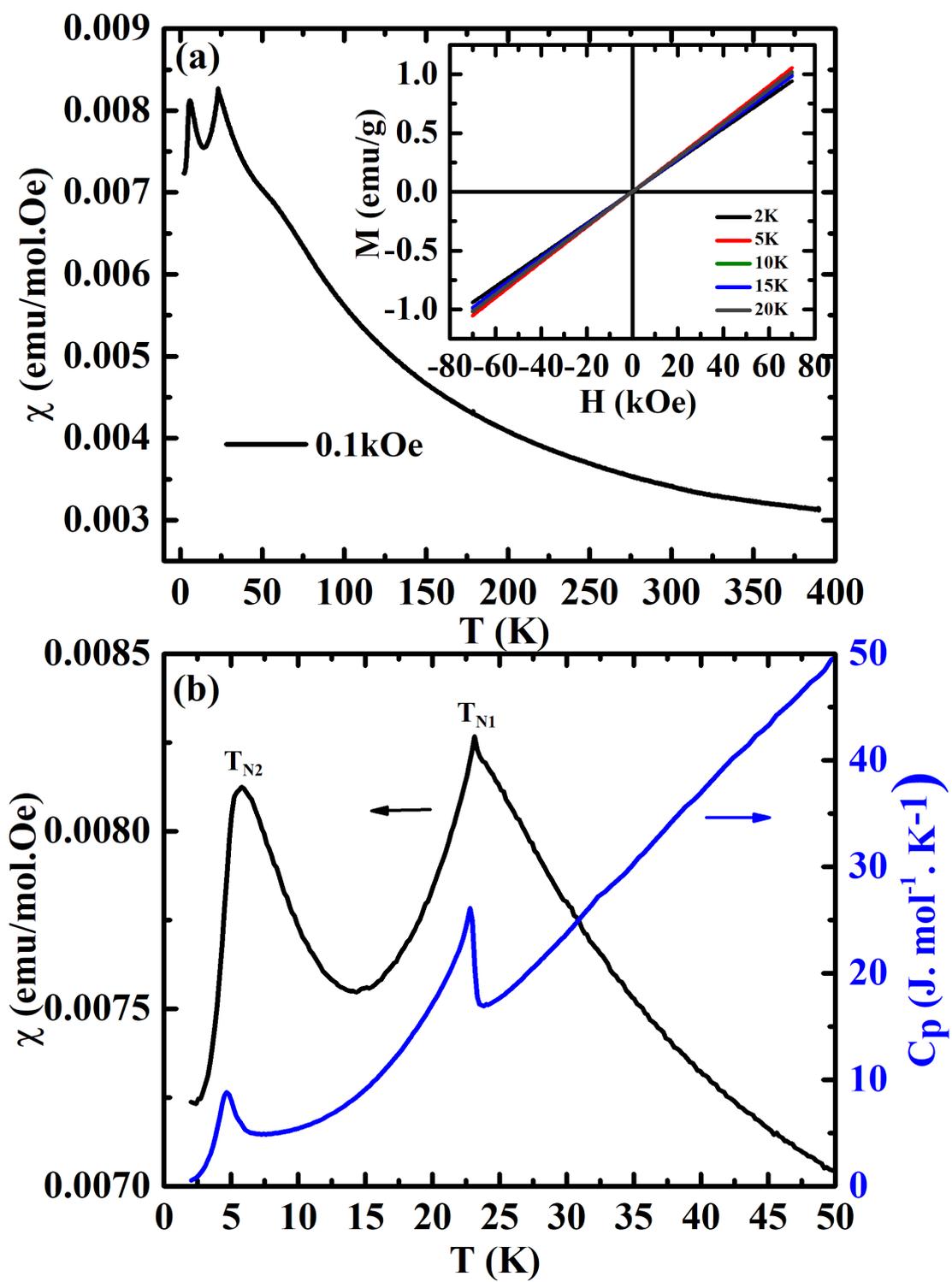

Figure 3:

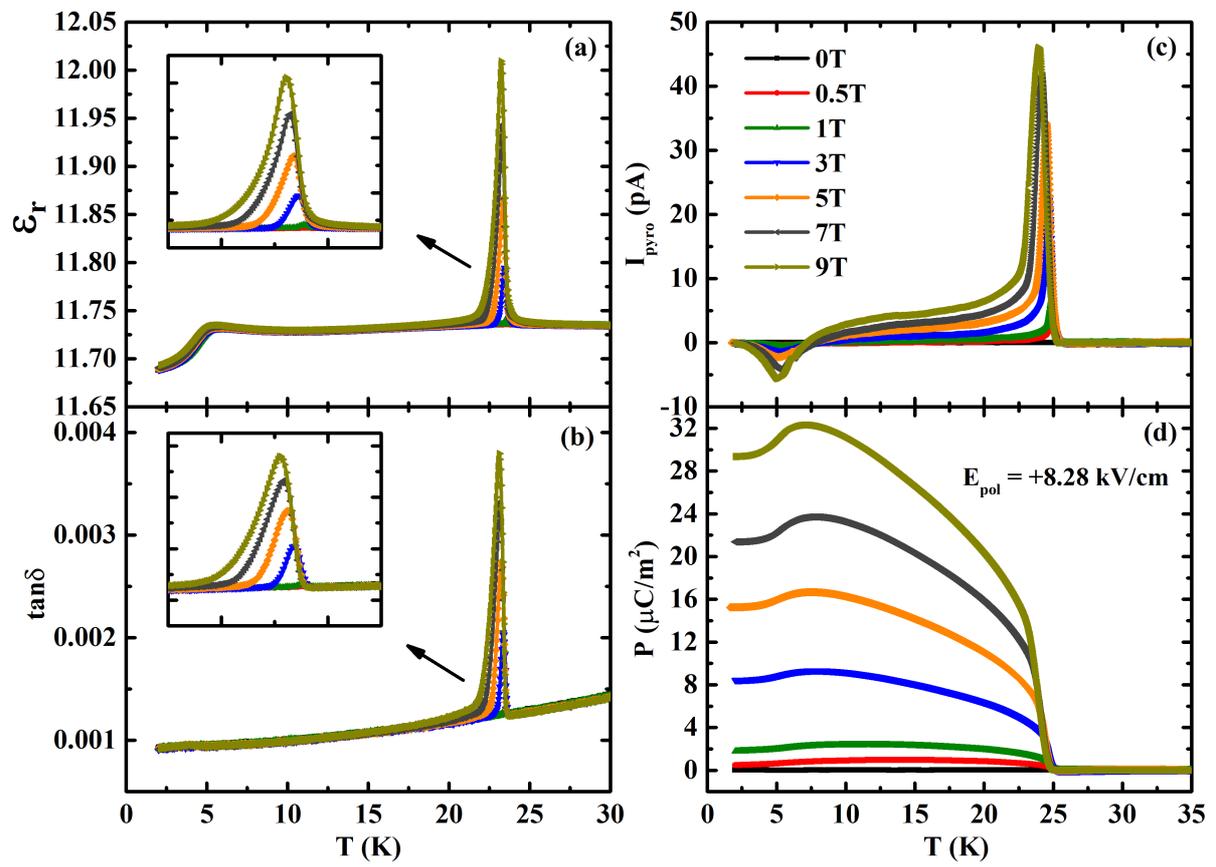

Figure 4:

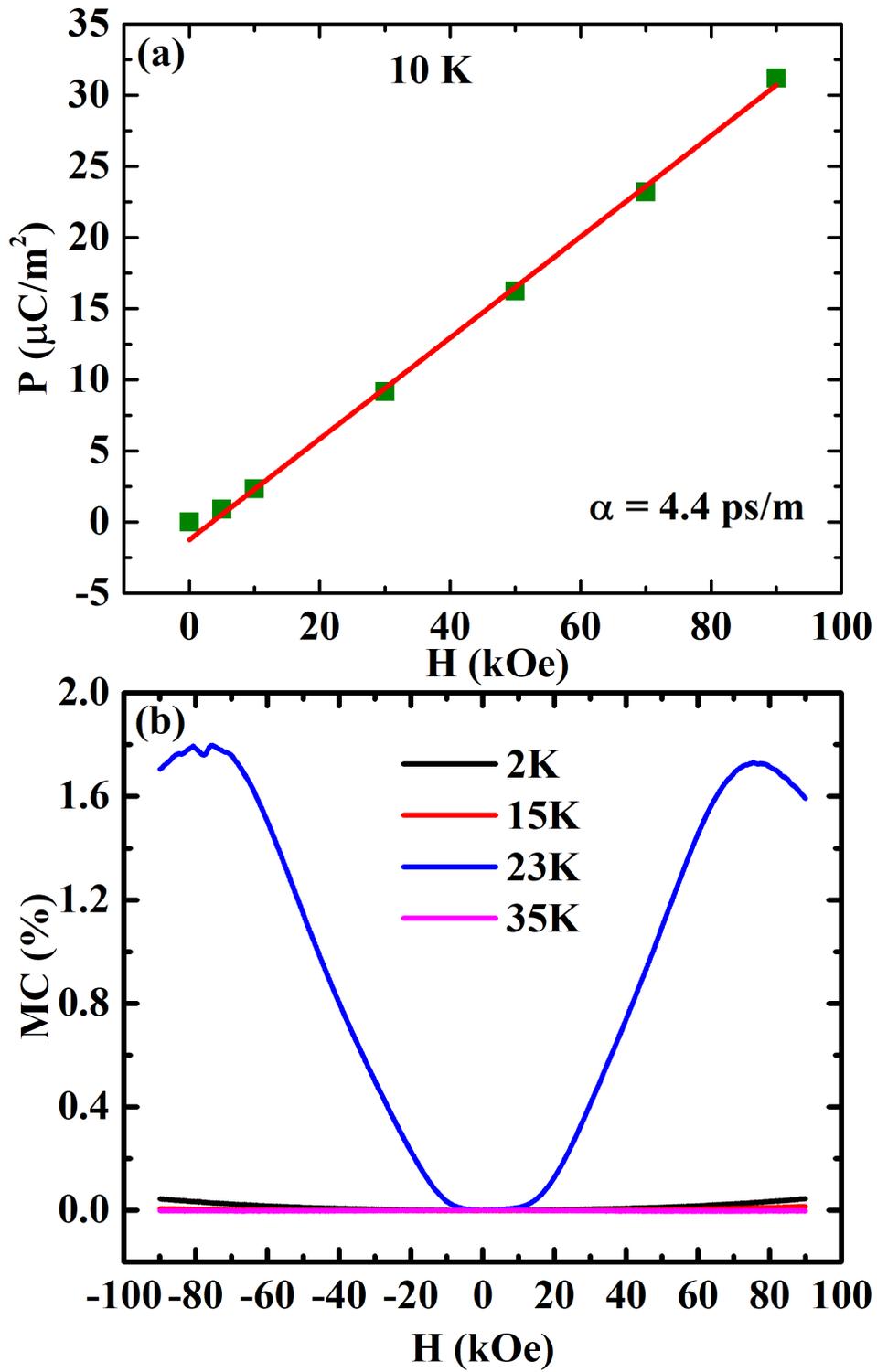

Figure 5:

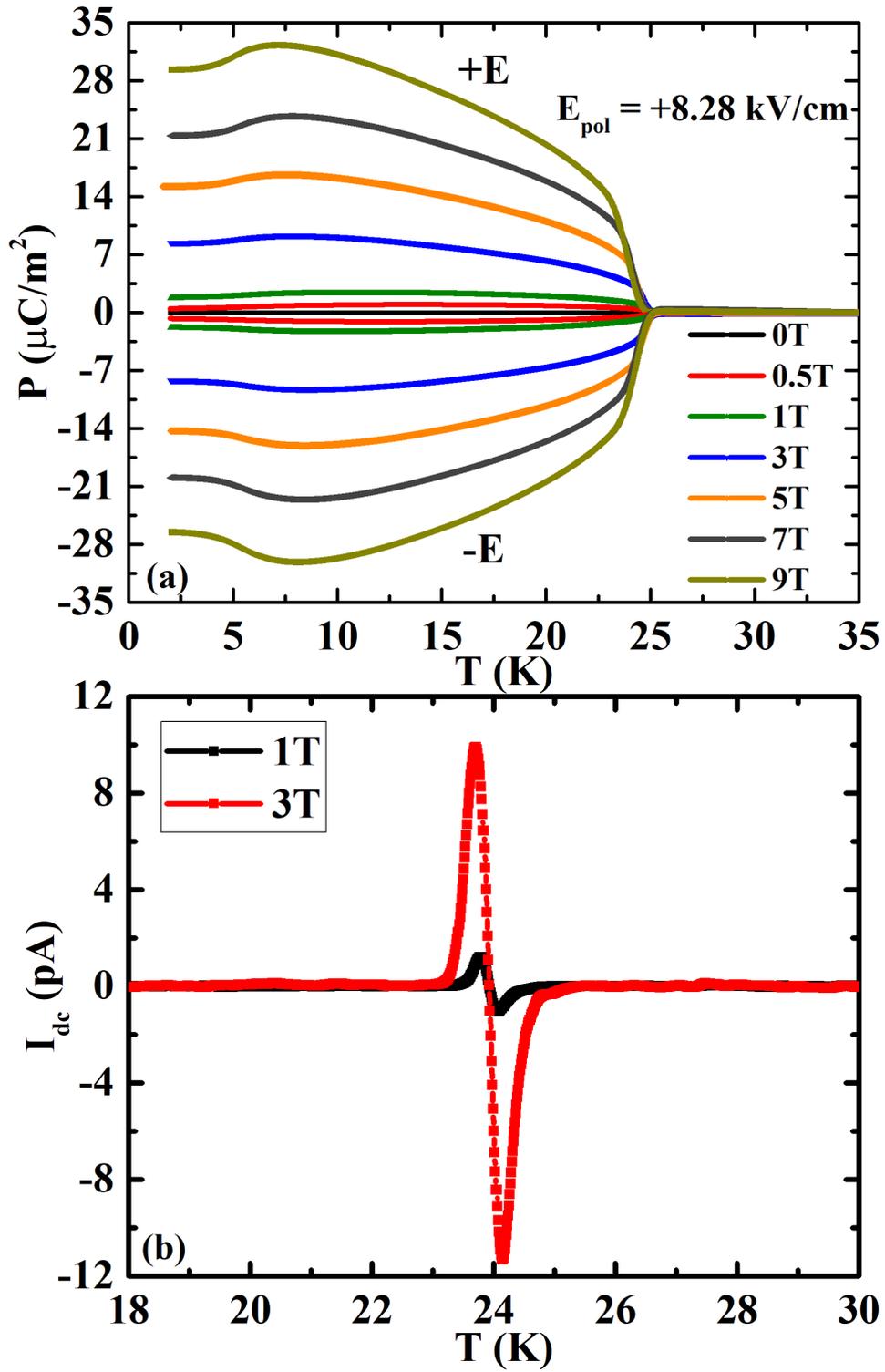